\documentclass[12pt]{article}
\usepackage{graphicx,epsf}

\newcommand{\MS}{$\overline{\mathrm{MS}}$\ }
\newcommand{\LMS}{\Lambda_{\overline{\mathrm{MS}}}}
\newcommand{\LQCD}{\Lambda_\mathrm{QCD}}
\newcommand{\ep}{\varepsilon}

\newcommand\cmp[3]{Comm.\ Math.\ Phys.\ {\bf #1} (#2) #3}

\newcommand\ijmpa[3]{Int.\ J.\ Mod.\ Phys.\ A {\bf #1} (#2) #3}
\newcommand\jhep[3]{J.\ High Ener.\ Phys.\ {\bf #1} (#2) #3}
\newcommand\npb[3]{Nucl.\ Phys.\ B {\bf #1} (#2) #3}
\newcommand\npps[3]{Nucl.\ Phys.\ B (Proc.\ Suppl.) {\bf #1} (#2) #3}
\newcommand\plb[3]{Phys.\ Lett.\ B {\bf #1} (#2) #3}
\newcommand\prd[3]{Phys.\ Rev.\ D {\bf #1} (#2) #3}
\newcommand\prep[3]{Phys.\ Rep.\ {\bf #1} (#2) #3}
\newcommand\prl[3]{Phys.\ Rev.\ Lett.\ {\bf #1} (#2) #3}

\newcommand\zpc[3]{Z.\ Phys.\ C {\bf #1} (#2) #3}
\newcommand{\hepph}[1]{{\tt hep-ph/#1}}

\begin{document}

\begin{titlepage}

\begin{flushright}
CLNS~01/1723\\
{\tt hep-ph/0102219}
\end{flushright}

\vspace{1.2cm}
\begin{center}
\Large\bf\boldmath
Asymptotics of QCD Factorization in Exclusive Hadronic Decays of $B$ 
Mesons
\unboldmath
\end{center}

\vspace{0.5cm}
\begin{center}
Thomas Becher, Matthias Neubert and Ben D. Pecjak\\[0.1cm]
{\sl Newman Laboratory of Nuclear Studies, Cornell University\\
Ithaca, NY 14853, USA}
\end{center}

\vspace{0.8cm}
\begin{abstract}
\vspace{0.2cm}\noindent
Using the renormalon calculus, we study the asymptotic behavior of
the perturbative expansion of the hard-scattering kernels entering
the QCD factorization formula for the nonleptonic weak decays
$\bar B^0\to D^{(*)+} M^-$, where $M$ is a light meson. In the 
``large-$\beta_0$ limit'', the kernels are infrared finite and free 
of endpoint singularities to all orders of perturbation theory. The 
leading infrared renormalon singularity corresponding to a power 
correction of order $\LQCD/m_b$ vanishes if the light meson has a 
symmetric light-cone distribution amplitude. We calculate the Borel 
transforms and the corresponding momentum distribution functions of 
the hard-scattering kernels, and resum the series of 
$O(\beta_0^{n-1}\alpha_s^n)$ corrections 
to explore the numerical significance of higher-order perturbative and 
power corrections. We also derive explicit expressions for the 
$O(\beta_0\,\alpha_s^2)$ contributions to the kernels, and for the 
renormalon singularities corresponding to power corrections of 
order $(\LQCD/m_b)^2$. Finally, we study the limit $m_c\to 0$ relevant 
to charmless hadronic decays such as $B\to\pi\pi$.
\end{abstract}

\vspace{1.0cm}
\centerline{\sl (Submitted to Nuclear Physics B)}

\vfill
\noindent
February 2001

\end{titlepage}

\section{Introduction}

The theoretical understanding of nonleptonic weak decays of hadrons
is complicated by the intricate effects of strong interactions. 
Gluon exchange between the quarks is characterized
by a multitude of relevant mass scales, ranging from the electroweak 
scale $\mu\sim M_W$ down to the confinement region $\mu\sim\LQCD$, 
where perturbative methods fail. Recently, however, it was 
shown that this picture simplifies drastically for most 
two-body hadronic decays of $B$ mesons into final states containing at 
least one fast, light meson \cite{BBNS,BBNS1}. In the heavy-quark limit,
the decay amplitudes for these processes factorize into a semileptonic
form factor and a meson decay constant. So-called ``nonfactorizable'' 
corrections are predominantly perturbative and taken into account by
convolutions of hard-scattering kernels with light-cone distribution
amplitudes of the mesons. Corrections to this limit are suppressed by 
powers of $\LQCD/m_b$.

Factorization as established in \cite{BBNS,BBNS1} is a 
nontrivial property of the decay amplitudes that holds true to all 
orders of perturbation theory and to leading power in $\LQCD/m_b$. 
For practical applications, it 
is important to obtain an estimate of the leading, 
power-suppressed corrections. Naive dimensional analysis suggests that 
such corrections are of order $\LQCD/m_b\approx 10\%$; however, it is 
difficult to quantify this statement. The main obstruction is that, in 
contrast to simpler applications of the heavy-quark expansion to 
exclusive semileptonic decays or inclusive processes \cite{review}, 
the power corrections to factorization cannot be organized using
a local operator product expansion. Whereas certain types of 
potentially large corrections were identified and estimated in 
\cite{BBNS}, it is not (yet) possible to write down the complete set 
of leading power corrections in terms of field-theoretic objects. 

The goal of the present work is to gain some insight into the structure
of power corrections arising from soft, ``nonfactorizable'' gluon 
exchange. We use the renormalon calculus \cite{Martin} to investigate 
the asymptotic behavior of the perturbation series for the 
hard-scattering kernels. These series are divergent and require
power corrections of a certain pattern in order to be consistently 
defined. The Borel transforms of the perturbation series have 
singularities in the complex plane, whose structure indicates the 
power (via their position) and strength (via their residues) of these
corrections. We study renormalon singularities by
adopting the ``large-$\beta_0$ limit'' 
(corresponding to the exchange of a single renormalon chain), in which 
nontrivial partial resummations of perturbation series become possible. 
This method provides the correct location of the renormalon 
singularities, but only approximately accounts for their residues.

To illustrate our approach, we recall the example
of the Adler $D$-function, i.e.\ the Euclidean correlator of two vector
currents. For this quantity, the leading infrared (IR) renormalon 
singularity arising 
from soft gluon exchange indicates a power correction of order 
$\LQCD^4/Q^4$, which has the same momentum dependence as the 
contribution of the gluon condensate. A consistent definition of the 
perturbation series with an accuracy of order $1/Q^4$ or better 
therefore requires inclusion of the gluon condensate, and the value of 
the condensate 
depends on the resummation prescription (e.g., principal-value Borel 
summation, truncation at the minimal term, etc.) adopted to define the 
divergent perturbation series \cite{Muel}. Although the presence 
of a $1/Q^4$ power correction proportional to the gluon condensate 
can be inferred from the operator product expansion, it 
is remarkable that the perturbation expansion itself signals the 
existence of this nonperturbative effect through its divergent 
large-order behavior. The correspondence of 
renormalon singularities and power corrections of (some) 
higher-dimensional operators in the operator product expansion has 
also been verified with several explicit examples in the context of 
the heavy-quark expansion \cite{BB,BSUV,NS}.

Building on this experience, the renormalon calculus has been applied to 
observables that do not admit an expansion in local operators. Examples 
are event-shape variables in $e^+ e^-$ annihilation 
\cite{MaWe95,Web94,Akh95,Nason,KoSt95}, Drell--Yan production 
\cite{KoSt95,Mar2}, fragmentation functions \cite{DaWe97,BBMa97}, and
structure functions in deep-inelastic scattering \cite{Stein,DaWe96}.
In these cases, there is no systematic framework that would allow 
us to classify the power corrections in terms of operator matrix
elements. However, the pattern of renormalon singularities determines 
at least a minimal set of power corrections that must be included for a 
consistent field-theoretic description. Although, in general, these are 
not the only sources of power-suppressed effects, the 
inclusion of the corrections corresponding to the leading IR renormalons 
significantly improves the phenomenological predictions. 
Interesting attempts to formalize this method include the dispersive
approach developed by Dokshitzer, Marchesini and Webber \cite{DMW96},
and the non-local operator method of Korchemsky and Sterman 
\cite{KoSt99}.

Our analysis will be similar in spirit to these approaches in that
we will use the renormalon calculus to obtain a minimal model of power
corrections to factorization in hadronic $B$ decays. 
We will also derive explicit results
for the presumably dominant part of the two-loop perturbative 
contributions to the hard-scattering kernels, i.e.\ the contributions
of order $\beta_0\,\alpha_s^2$. For simplicity, we focus mainly 
on the class-1 $B$ decays into a heavy--light final state such 
as $\bar B^0\to D^{(*)+} M^-$, where $M=\pi,\rho,\dots$ is a light 
meson. For these processes, the factorization formula takes its simplest 
form and is best established theoretically. 

The effective weak Hamiltonian is (considering 
$b\to c\bar u d$ transitions for concreteness)
\begin{equation}
   {\cal H}_{\rm eff} = \frac{G_F}{\sqrt2}\,V_{cb} V_{ud}^*\,
   \Big[ C_1(\mu) O_1(\mu) + C_2(\mu) O_2(\mu) \Big] \,,
\end{equation}
where
\begin{eqnarray}\label{ops}
   O_1 &=& \bar d_\alpha\gamma_\mu(1-\gamma_5)u_\alpha\,
         \bar c_\beta\gamma^\mu(1-\gamma_5)b_\beta \,, \nonumber\\
   O_2 &=& \bar d_\alpha\gamma_\mu(1-\gamma_5)u_\beta\,
         \bar c_\beta\gamma^\mu(1-\gamma_5)b_\alpha
\end{eqnarray}
are local four-quark operators, summed over color indices $\alpha,\beta$. 
The Wilson coefficients $C_i(\mu)$ contain short-distance corrections 
arising from gluon exchange with virtualities between the electroweak 
scale $M_W$ and the renormalization scale $\mu\sim m_b$. 
The hadronic matrix elements of the operators $O_i(\mu)$ greatly 
simplify in the heavy-quark limit $m_b\gg\LQCD$. To leading power in 
$\LQCD/m_b$, and to all orders of perturbation theory, they can be 
expressed in the factorized form \cite{BBNS}
\begin{eqnarray}\label{fact}
   &&\hspace{-0.6cm}
    \langle M^- D^{(*)+}|\,O_i(\mu)\,|\bar B^0\rangle \nonumber\\
   &=& \int\limits_0^1\!dx\,\Phi_M(x,\mu)\,\bigg[ T_{iL}(x,\mu)\,
    \langle M^-|\,\bar d\gamma_\mu(1-\gamma_5)u\,|\,0\,\rangle
    \langle D^{(*)+}|\,\bar c\gamma^\mu(1-\gamma_5)b\,|\bar B^0\rangle
    \nonumber\\[-0.2cm]
   &&\hspace{2.22cm} \mbox{}+ T_{iR}(x,\mu)\,
    \langle M^-|\,\bar d\gamma_\mu(1-\gamma_5)u\,|\,0\,\rangle
    \langle D^{(*)+}|\,\bar c\gamma^\mu(1+\gamma_5)b\,|\bar B^0\rangle
    \bigg] \,.
\end{eqnarray}
$\Phi_M(x,\mu)$ is the leading-twist light-cone distribution amplitude 
of the light meson $M$, normalized such that $\int_0^1 dx\,
\Phi_M(x,\mu)=1$, and $x$ is the longitudinal momentum fraction of the 
$d$ quark in the meson. The current matrix elements on the right-hand 
side of the factorization formula can be expressed in terms of 
$\bar B\to D^{(*)}$ semileptonic form factors and the decay constant
$f_M$ of the meson $M$. In the heavy-quark limit, 
``nonfactorizable'' strong-interaction effects are 
dominated by hard gluon exchange and included in the hard-scattering
kernels $T_i(x,\mu)$. In \cite{BBNS}, these kernels were calculated 
explicitly at one-loop order and were shown to be free of 
IR divergences at the two-loop order.

The goal of the present work is to estimate the higher-order corrections 
in the loop expansion of the hard-scattering kernels and, at the same 
time, to gain some insight into 
the structure of power corrections to factorization. In 
Section~\ref{sec:Borel} we derive expressions for the Borel transforms 
of the hard-scattering 
kernels using the single renormalon-chain approximation. We 
present resummation formulae for the kernels in the large-$\beta_0$
limit and establish several important results: the systematic
cancellation of IR divergences to all orders of perturbation theory, the
absence of endpoint singularities in the kernels, and the vanishing of
the leading renormalon singularity (corresponding to a first-order power
correction) in the case of a symmetric light-cone
distribution amplitude. In Section~\ref{sec:pert} we study higher-order 
contributions in the perturbative expansion of the kernels. We derive 
exact results for the terms of order $\beta_0\,\alpha_s^2$, and give 
all-order expressions for the anomalous dimensions of the four-quark
operators in the large-$\beta_0$ limit. We also investigate numerically
the asymptotic behavior of the perturbation series for the 
hard-scattering kernels. A systematic study of IR renormalons and power
corrections is performed in Section~\ref{sec:power}. We present explicit
results for the first two IR renormalon singularities corresponding to
power corrections of order $\LQCD/m_b$ and $(\LQCD/m_b)^2$. In 
Section~\ref{sec:zto0} we study the limit $m_c\to 0$ and derive results
valid for massless quarks in the final-state. They are relevant to 
charmless decays such as $B\to\pi\pi$.

\section{Borel transforms and distribution functions}
\label{sec:Borel}

The perturbation series for the hard-scattering kernels $T_i(x,\mu)$ in
the factorization formula (\ref{fact}) can be arranged as
\begin{equation}\label{Series}
   T_i(x,\mu) = t_i^{(0)}
   + \sum_{\ell=1}^\infty \sum_{n=0}^{\ell-1}\,
   t_i^{(n,\ell)}(x,\mu)\,\beta_0^n\alpha_s^\ell(\mu) \,,
\end{equation}
where $\ell$ is the number of loops, and 
$\beta_0=\frac{11}{3}\,C_A - \frac{4}{3}\,T_F n_f$ is the first 
coefficient of the $\beta$-function for an $\mbox{SU}(N_c)$ gauge 
theory ($C_A=N_c$ and $T_F=\frac12$) with $n_f$ light quark flavors.
The tree-level coefficients are $t_{1L}^{(0)}=1$, $t_{2L}^{(0)}=1/N_c$, 
and $t_{iR}^{(0)}=0$.
Our goal is to sum the terms of order $\beta_0^{\ell-1}\alpha_s^\ell$ 
to all orders of perturbation theory. In other words, we consider the 
limit of large $\beta_0$ for fixed $\beta_0\,\alpha_s$ and calculate the 
kernels $T_i(x,\mu)$ to order $1/\beta_0$, neglecting terms of order 
$1/\beta_0^2$ and higher. Strictly speaking, there is no sensible 
limit of QCD in which $\beta_0$ may be considered a large parameter 
(in particular, this is not the large-$N_c$ limit) -- except, perhaps, 
taking $n_f\to-\infty$. Nevertheless, retaining only the leading terms 
in $1/\beta_0$ often gives a good approximation to exact multi-loop 
results (see, e.g., \cite{BG}), in particular in cases when there is 
a nearby IR renormalon \cite{scale}.

\boldmath
\subsection{Borel summation in the large-$\beta_0$ limit}
\unboldmath

The coefficients $t_i^{(\ell-1,\ell)}$ of the terms with the highest 
degree of $\beta_0$ in (\ref{Series}) are determined by diagrams with 
$(\ell-1)$ light-quark loops, which are rather straightforward to 
calculate. We work in dimensional regularization with $4-2\ep$ 
space-time dimensions and adopt the \MS subtraction scheme. At first 
order in $1/\beta_0$, coupling-constant
renormalization is accomplished by ($\bar\mu^2=\mu^2 e^\gamma/4\pi$)
\begin{equation}
   \frac{\beta_0\,g_s^2}{(4\pi)^2}
    = \frac{\bar\mu^{2\ep}\,b(\mu)}{1+b(\mu)/\ep} \,, \qquad
   b(\mu) = \frac{\beta_0\,\alpha_s(\mu^2)}{4\pi}
    = \frac{1}{\ln(\mu^2/\LMS^2)} \,.
\end{equation}
To leading order in the large-$\beta_0$ limit the ``nonfactorizable'' 
vertex corrections to the operator $O_1$ in (\ref{ops}) vanish when
projected onto color-singlet meson states, i.e.\
\begin{equation}
   T_{1L}(x,\mu) = 1 + O(1/\beta_0^2) \,, \qquad
   T_{1R}(x,\mu) = O(1/\beta_0^2) \,. 
\end{equation}
The perturbation series for the kernels corresponding to the operator 
$O_2$ can be written as
\begin{eqnarray}\label{Struct}
   T_{2L}(x,\mu) &=& \frac{1}{N_c} 
    \left[ \,1 + \frac{2C_F}{\beta_0}\,\sum_{\ell=1}^{\infty}\,
    \frac{F_L(\ep,\ell\ep)}{\ell} \bigg( \frac{b}{\ep+b} \bigg)^\ell
    - \hbox{(UV subtractions)} + O(1/\beta_0^2) \right] , \nonumber\\
   T_{2R}(x,\mu) &=& \frac{1}{N_c} 
    \left[ \,\frac{2C_F}{\beta_0}\,\sum_{\ell=1}^{\infty}\,
    \frac{F_R(\ep,\ell\ep)}{\ell} \bigg( \frac{b}{\ep+b} \bigg)^\ell
    + O(1/\beta_0^2) \right] ,
\end{eqnarray}
where $b=b(\mu)$. The kernel $T_{2R}$ corresponding to the right-handed 
(i.e.\ chirality-flipped) heavy-quark current in (\ref{fact}) is 
ultraviolet (UV) finite in the large-$\beta_0$ limit and does not 
require \MS subtractions. 

The functions $F_i(\ep,u)$ are regular at $\ep=u=0$. Expanding 
$F_i(\ep,u)$ in powers of $\ep$ and 
$u$, and $[b/(\ep+b)]^\ell$ in powers of $b/\ep$, gives a quadruple 
sum in (\ref{Struct}). Combinatoric identities relate the $1/\ep$ terms, 
and hence the anomalous dimensions of the local four-quark operators 
$O_1$ and $O_2$, to the Taylor coefficients of $F_L(\ep,0)$ \cite{meth1}.
Using the well-known fact that the linear combinations $O_\pm=O_1\pm O_2$
are multiplicatively renormalized, we find that the corresponding 
anomalous dimensions $\gamma_\pm$ are given by
\begin{equation}\label{gamma}
   \gamma_\pm = \pm \frac{N_c\mp 1}{N_c}\,\frac{(-2b)}{\beta_0}\,
   F_L(-b,0) + O(1/\beta_0^2) \,.
\end{equation}
The finite terms, which determine the kernels themselves, receive 
contributions from the Taylor coefficients of $F_L(\ep,0)$ and 
$F_i(0,u)$ \cite{meth2}. We obtain
\begin{eqnarray}\label{finite}
   N_c\,T_{2L}(x,\mu) &=& 1 + \frac{2C_F}{\beta_0} 
    \int\limits_{-b(\mu)}^0\!d\ep\,\frac{F_L(\ep,0)-F_L(0,0)}{\ep}
    \nonumber\\
   &&\mbox{}+ \frac{2C_F}{\beta_0} \int\limits_0^\infty\!du\,
    e^{-u/b(\mu)}\,\frac{F_L(0,u)-F_L(0,0)}{u} + O(1/\beta_0^2) \,,
    \nonumber\\
   N_c\,T_{2R}(x,\mu) &=& \frac{2C_F}{\beta_0} \int\limits_0^\infty\!du\,
    e^{-u/b(\mu)}\,\frac{F_R(0,u)}{u} + O(1/\beta_0^2) \,.
\end{eqnarray}
If they were unambiguously defined, these expressions would determine
the Borel-resum\-med perturbation series for the hard-scattering kernels 
in the large-$\beta_0$ limit. However, the presence 
of singularities located on the integration contour renders the 
integrals over $u$ ambiguous. These (renormalon) singularities
will provide us with information about power corrections to
the factorization formula (\ref{fact}).

\begin{figure}
\centerline{\epsfxsize=14cm\epsffile{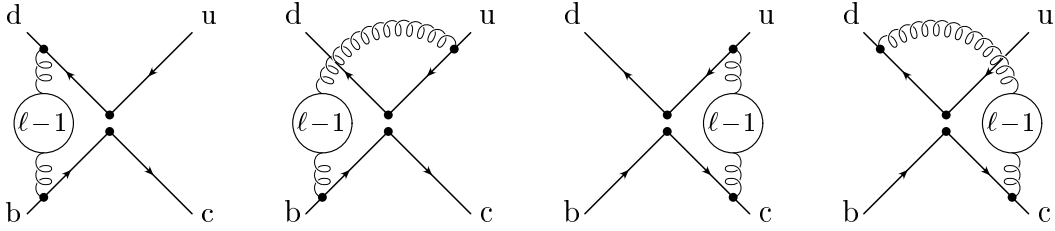}}
\centerline{\parbox{14cm}{\caption{\label{fig:graphs}
``Nonfactorizable'' vertex diagrams contributing to the hard-scattering
kernels in the factorization formula. The split vertex represents an
insertion of one of the four-quark operators $O_i$. The circles 
represent $(\ell-1)$ insertions of light-quark loops on the gluon 
propagators.}}}
\end{figure}

The functions $F_L(\ep,u)$ and $F_R(\ep,u)$ are obtained by computing
the ``nonfactorizable'' vertex corrections shown in 
Figure~\ref{fig:graphs}. It is convenient to introduce the abbreviations
\begin{equation}\label{deltadef}
   z = \frac{m_c}{m_b} \,, \qquad
   \delta = (1-z^2)\,x \,, \qquad
   \bar\delta = (1-z^2)(1-x) \,.
\end{equation}
The variable $\delta$ ($\bar\delta$) appears in the calculation of
the first (second) diagram in Figure~\ref{fig:graphs}. The third and 
fourth diagrams are obtained using crossing symmetry and require the 
substitutions
\begin{equation}\label{crossing}
   m_b\to m_c \,, ~\quad z\to\frac{1}{z} \,, ~\quad x\to (1-x) \,,
   ~\quad \delta\to - \frac{\bar\delta}{z^2} - i\epsilon \,, ~\quad
   \bar\delta\to - \frac{\delta}{z^2} - i\epsilon \,.
\end{equation}
These contributions, to which we will refer as ``crossed terms'', 
contain imaginary parts (specified by the $i\epsilon$ 
prescriptions), which determine the strong-interaction phases 
of the decay amplitudes. From now on we will omit the $i\epsilon$'s; 
they can be reinstated by recalling that $z^2\equiv z^2-i\epsilon$.
The results are
\begin{eqnarray}\label{Fres}
   F_L(\ep,u) &=& \left( \frac{\mu}{m_b} \right)^{2u} e^{\gamma\ep}\,
    [D(\ep)]^{u/\ep-1}\,\Big[ A_1(\delta,\ep,u) - A_2(\bar\delta,\ep,u)
    \Big] + \mbox{crossed terms} \,, \nonumber\\
   F_R(\ep,u) &=& \left( \frac{\mu}{m_b} \right)^{2u} e^{\gamma\ep}\,
    [D(\ep)]^{u/\ep-1}\,z\,B_1(\delta,\ep,u) + \mbox{crossed terms} \,,
\end{eqnarray}
where
\begin{eqnarray}\label{Ai}
   A_1(r,\ep,u) &=& (1-\ep)^2\,(r\,g_{11}+g_{20})
    + u\,(g_{10}-g_{20}) \nonumber\\
   &&\mbox{}- u r\,(g_{00}-g_{01}-g_{10}+g_{11}) \,, \nonumber\\
   A_2(r,\ep,u) &=& (4-\ep-\ep^2)(r\,g_{11}+g_{20})
    + u\,(g_{10}-2 g_{20}) \nonumber\\
   &&\mbox{}- u r\,(g_{00}-g_{01}-g_{10}+2 g_{11}) \,, \nonumber\\
   B_1(r,\ep,u) &=& -u\,g_{20} \,,
\end{eqnarray}
and
\begin{eqnarray}\label{gmn}
   g_{mn} \equiv g_{mn}(r,\ep,u)
   = \frac{\Gamma(m+n-2u)\,\Gamma(1+u)}{\Gamma(1+m+n-u-\ep)}\,
   \int\limits_0^1\!dy\,
   \frac{y^m(1-y)^n}{\big[y(r+(1-r)y)\big]^{1+u}} \,.
\end{eqnarray}
The integral can be expressed in terms of incomplete Euler 
$\beta$-functions. The notation in (\ref{Fres}) is such that the 
functions $A_1$ and $B_1$ correspond to the contributions of the first 
diagram, whereas $A_2$ is obtained from the second diagram. The linear 
factor of $z$ in front of $B_1$ comes from the chirality flip on the 
charm-quark line. The 
contributions from the third and fourth diagrams are obtained by 
performing the substitutions shown in (\ref{crossing}). Finally, the 
function
\begin{equation}\label{De}
   D(\ep) = 6 e^{\gamma\ep}\,
   \frac{\Gamma(1+\ep)\,\Gamma^2(2-\ep)}{\Gamma(4-2\ep)} 
   = 1 + \frac53\,\ep + O(\ep^2) 
\end{equation}
is related to the contribution of a light-quark loop to the gluon
self-energy. 

In the limit $u\to 0$ most of the terms in (\ref{Ai}) vanish. However, 
the functions $g_{00}$ and $g_{01}$ have poles at $u=0$, corresponding 
to soft and collinear IR singularities of the individual diagrams.
Working out the residues we find
\begin{equation}\label{F12}
   F_L(\ep,0) = e^{\gamma\ep}\,[D(\ep)]^{-1}\,\left[ 
   - \frac{(3-2\ep)(1+\ep)}{\Gamma(3-\ep)}
   + \frac{1}{\Gamma(1-\ep)}\,\ln\frac{x}{1-x} \right]
   + \mbox{crossed terms} \,.
\end{equation}
The logarithm corresponds to a remaining collinear singularity in the 
sum of the first two diagrams, and cancels when adding the crossed 
terms \cite{BBNS}. As a consequence, only the first term in the square 
brackets, which comes from 
the first term in the expressions for $A_1$ and $A_2$ in (\ref{Ai}), 
remains. This term depends on the renormalization scheme. 
Specifically, the $\ep$-dependent factors in (\ref{Ai}) follow
from the Dirac identities
\begin{eqnarray}
    \gamma_\mu\gamma_\nu\gamma_\alpha(1-\gamma_5)\otimes
    \gamma^\alpha\gamma^\nu\gamma^\mu(1-\gamma_5)
    &=& 4(1-\ep)^2\,
     \gamma_\rho(1-\gamma_5)\otimes\gamma^\rho(1-\gamma_5) \,,
     \nonumber\\
    \gamma_\mu\gamma_\nu\gamma_\alpha(1-\gamma_5)\otimes
    \gamma^\mu\gamma^\nu\gamma^\alpha(1-\gamma_5)
    &=& 4(4-\ep-\ep^2)\,
     \gamma_\rho(1-\gamma_5)\otimes\gamma^\rho(1-\gamma_5) \,,
\end{eqnarray}
which are valid in the so-called ``naive dimensional regularization'' 
(NDR) scheme with anticommuting $\gamma_5$, and with a projection of
evanescent operators as specified in eq.~(4.3) of \cite{BJLW}. The 
scheme-dependence cancels against that of the Wilson coefficients in 
the effective weak Hamiltonian. Using the expression for 
$D(\ep)$ from (\ref{De}), we obtain
\begin{eqnarray}\label{FLep0}
   F_L(\ep,0) &=& - \frac{(3-2\ep)(1+\ep)\,\Gamma(4-2\ep)}
    {3\Gamma(1+\ep)\,\Gamma^2(2-\ep)\,\Gamma(3-\ep)} \,, \qquad
   F_L(0,0) = - 3 \,, 
\end{eqnarray}
and $F_R(\ep,0)=0$. In the opposite limit $\ep\to 0$ we find
\begin{eqnarray}\label{Fi0u}
   F_L(0,u) &=& \left( \frac{\mu}{m_b} \right)^{2u} e^{5u/3}\,
    \Big[ A_1(\delta,0,u) - A_2(\bar\delta,0,u) \Big]
    + \mbox{crossed terms} \,, \nonumber\\
   F_R(0,u) &=& \left( \frac{\mu}{m_b} \right)^{2u} e^{5u/3}\,
    z\,B_1(\delta,0,u) + \mbox{crossed term} \,.
\end{eqnarray}

We now rewrite the resummed expressions (\ref{finite}) for the kernels 
in a more convenient form. The $\mu$-dependent factor in the 
expressions for $F_i(0,u)$ in (\ref{Fi0u}) combines with the 
exponential $e^{-u/b(\mu)}$ in (\ref{finite}) into the 
renormalization-scheme invariant combination
\begin{equation}
   e^{-u/b(\mu)} \left( \frac{\mu}{m_b} \right)^{2u} e^{5u/3}
   = \left( \frac{\LMS}{m_b} \right)^{2u} e^{5u/3}
   \equiv \left( \frac{\Lambda_{\rm V}}{m_b} \right)^{2u}
   = e^{-u/b_V(m_b)} \,,
\end{equation}
where $\Lambda_{\rm V}=e^{5/6}\LMS$ is the QCD scale parameter in the 
so-called ``V scheme'' \cite{BLM}, and 
\begin{equation}\label{bVdef}
   b_V(m_b) = \frac{1}{\ln(m_b^2/\Lambda_{\rm V}^2)}
   = \frac{\beta_0\,\alpha_s^{(V)}(m_b^2)}{4\pi}
   = \frac{\beta_0\,\alpha_s(e^{-5/3} m_b^2)}{4\pi}
\end{equation}
is proportional the running coupling constant in that scheme. Here, as 
always, $\alpha_s(\mu^2)$ without a label ``V'' is the coupling 
constant in the \MS scheme. We define the Borel transform of the 
hard-scattering kernel $T_{2R}$ as
\begin{equation}
   S_R(u,x) \equiv e^{-5u/3}\,\frac{F_R(0,u)}{u} \bigg|_{\mu=m_b}
   = z\,\frac{B_1(\delta,0,u)}{u}
   + z^{-1-2u}\,\frac{B_1(-\bar\delta/z^2,0,u)}{u} \,.
\end{equation}
The factor $z^{-2u}$ in front of the crossed term comes from the 
fact that the scale factor associated with this term is
$(\mu/m_c)^{2u}=z^{-2u}\,(\mu/m_b)^{2u}$. With this definition, the 
Borel integral representation for the kernel becomes
\begin{equation}\label{T1Rfinal}
   N_c\,T_{2R}(x,\mu) = \frac{2 C_F}{\beta_0}
   \int\limits_0^\infty\!du\,e^{-u/b_V(m_b)}\,S_R(u,x)
   + O(1/\beta_0^2) \,.
\end{equation}
For the kernel $T_{2L}$, we define the Borel transform as
\begin{eqnarray}
   S_L(u,x) &\equiv& \frac{e^{-5u/3}}{u} \left[
    F_L(0,u) \Big|_{\mu=m_b} - F_L^{(12)}(0,0)
    - z^{-2u}\,F_L^{(34)}(0,0) \right] \\
   &=& \frac{1}{u} \bigg[ A_1(\delta,0,u) -\! A_2(\bar\delta,0,u)
    + e^{-5u/3} \left( \frac32 - \ln\frac{x}{1-x} \right) 
    + z^{-2u}\!\times\!\mbox{(crossed terms)} \bigg] \,. \nonumber
\end{eqnarray}
Here $F_L^{(12)}(0,0)=-\frac32+\ln\frac{x}{1-x}$ and 
$F_L^{(34)}(0,0)=-\frac32-\ln\frac{x}{1-x}$ correspond to the 
contributions of the only first two and the last two diagrams in 
Figure~\ref{fig:graphs}, respectively, as can be seen by taking the 
limit $\ep\to 0$ in (\ref{F12}). With the above definition, the two 
parts of the Borel transform corresponding to each pair of diagrams are
separately free of UV and IR singularities at $u=0$, and the result for
the kernel takes the form
\begin{eqnarray}\label{T1Lfinal}
   N_c\,T_{2L}(x,\mu) &=& 1 + \frac{2 C_F}{\beta_0}
    \int\limits_{-b(\mu)}^0 \frac{d\ep}{\ep}\,\left[ 
    3 - \frac{(3-2\ep)(1+\ep)\Gamma(4-2\ep)}
             {3\Gamma(1+\ep)\,\Gamma^2(2-\ep)\,\Gamma(3-\ep)} \right]
    \nonumber\\
   &&+ \frac{2 C_F}{\beta_0} \left[ 3 \ln\frac{b(\mu)}{b(m_b)} 
    - \left( \frac32 + \ln\frac{x}{1-x} \right) 
    \ln\frac{b(m_c)}{b(m_b)} \right] \nonumber\\
   &&+ \frac{2 C_F}{\beta_0} \int\limits_0^\infty\!du\,
    e^{-u/b_V(m_b)}\,S_L(u,x) + O(1/\beta_0^2) \,.
\end{eqnarray}

In the above expressions for the hard-scattering kernels, the Borel 
integrals (the integrals over $u$) contain all 
nontrivial information about the asymptotic divergence of the 
perturbation expansions. They are defined in a renormalization-scheme 
invariant way. So-called IR
renormalon singularities located on the integration contour along the 
positive real $u$-axis render these integrals ill-defined, and hence
the perturbation series are not Borel summable \cite{reno1,reno2,tHof}. 
In the large-$\beta_0$ limit, renormalon singularities show up as poles 
at half-integer values of $u$, as seen from the explicit form of
the functions $g_{mn}$ in (\ref{gmn}). A pole singularity located at 
$u=k/2$ corresponds to a factorial growth $t_i^{(\ell-1,\ell)}\sim 
(k/2)^{-\ell}\,(\ell-1)!$ of the perturbative expansion coefficients 
in (\ref{Series}), and leads to an irreducible ambiguity of order
$(\LQCD/m_b)^k$ in the definition of the resummed series. From the
expression for $g_{mn}$ it follows that for $n=0$ these functions have 
single poles at $u=\frac12$. Working out the residues we obtain
\begin{equation}\label{pole12}
   S_L(u,x) \stackrel{u\to\frac12}{=} - \frac{4}{1-z}
   \left( \frac{1}{x} - \frac{1}{1-x} \right) \frac{1}{1-2u}
   + \mbox{regular terms} \,,
\end{equation}
whereas $S_R(u,x)$ is regular at $u=\frac12$. A pole at $u=\frac12$
corresponds to a power correction of first order in $\LQCD/m_b$, which 
could be large for realistic heavy-quark masses. However, because the 
residue of the pole in (\ref{pole12}) is
antisymmetric under exchange of $x\leftrightarrow(1-x)$, this 
contribution vanishes for the case of a symmetric light-cone 
distribution amplitude. Therefore, the renormalon analysis does not
indicate a first-order power correction for decays such as
$\bar B^0\to D^{(*)+}\pi^-$ or $\bar B^0\to D^{(*)+}\rho^-$. A more 
detailed investigation of 
power corrections will be presented in Section~\ref{sec:power}.

\subsection{Momentum distribution functions}

The analysis of higher-order renormalon singularities and the numerical 
evaluation of the Borel integrals in the form 
of (\ref{T1Rfinal}) and (\ref{T1Lfinal}) are difficult because of the
complicated structure of the Borel transforms $S_i(u,x)$. It is 
advantageous to rewrite the Borel integrals as integrals over a running 
coupling constant multiplied by weight functions, using the formalism 
developed in \cite{scale}:
\begin{equation}\label{distr}
   \frac{1}{\beta_0} \int\limits_0^\infty\!du\,e^{-u/b_V(m_b)}\,S_i(u,x)
   = \int\limits_0^\infty\frac{d\tau}{\tau}\,
   w_i(\tau,x)\,\frac{\alpha_s(\tau\,e^{-5/3}\,m_b^2)}{4\pi} \,.
\end{equation}
This elucidates the connection between IR (UV) renormalons
and small-momentum (large-momentum) contributions in Feynman diagrams. 
The functions $w_i(\tau,x)$ determine the distribution of 
the gluon virtualities inside the vertex-correction diagrams in 
Figure~\ref{fig:graphs}. Technically, they are the inverse Mellin 
transforms of the Borel images $S_i(u,x)$. The running coupling constant 
under the integral on the right-hand side is nothing but the coupling 
$\alpha_s^{(V)}(\tau\,m_b^2)$ in the V scheme, as defined in
(\ref{bVdef}). Equation~(\ref{distr}) has been established in
the literature as an integral representation of the Borel sum for
quantities defined in Euclidean kinematics \cite{scale}. It is applied 
here for the first time to a situation in which the quantity of interest 
has a more complicated analytic structure and a nonvanishing dispersive 
part. We have checked  numerically that the two sides in (\ref{distr})
agree even if the imaginary part is nonzero.

The distribution functions $w_i(\tau,x)$ can be computed in terms of 
Feynman parameter integrals starting from the relation \cite{scale}
\begin{equation}
   S_i(u,x) = \int\limits_0^\infty\frac{d\tau}{\tau}\,w_i(\tau,x)\,
   \tau^{-u} \,,
\end{equation}
which is valid for $-1<\mbox{Re}\,u<\frac12$. The result is
\begin{eqnarray}\label{wLwR}
   w_L(\tau,x) &=& f_L(\tau,\delta,\bar\delta)
    + f_L(\tau/z^2,-\bar\delta/z^2,-\delta/z^2)
    \,, \nonumber\\
   w_R(\tau,x) &=& z\,f_R(\tau,\delta)
    + \frac{1}{z}\,f_R(\tau/z^2,-\bar\delta/z^2) \,,
\end{eqnarray}
where the first (second) terms on the right-hand side correspond to the 
first (last) two diagrams in Figure~\ref{fig:graphs}. We find
\begin{eqnarray}\label{fLRres}
   f_L(\tau,\delta,\bar\delta)
   &=& \tau \bigg( \frac{1+\eta}{2\delta} - \frac{\eta}{\delta^2} \bigg)
    - \frac{\tau}{\delta}
    \bigg( 2 - \tau\,\frac{1-\delta}{\delta^2} \bigg)  
    \ln\bigg( 1 + \frac{\eta\,\delta}{\tau} \bigg)
    + \frac32\,\theta(\tau\,e^{-5/3} - 1) \nonumber\\
   &&\mbox{}+ \left[ - \frac{\eta}{\delta}
    - \bigg( 1 - \frac{\tau}{\delta^2} \bigg)
    \ln\bigg( 1 + \frac{\eta\,\delta}{\tau} \bigg)
    + \ln\delta\cdot\theta(1 - \tau\,e^{-5/3})
    - \{ \delta\to\bar\delta \} \right] , \nonumber\\ 
   f_R(\tau,\delta) &=& \tau
    \left[ - \frac{1-\eta}{2\delta} + \frac{\eta}{\delta^2}
    - \frac{\tau}{\delta^3}
    \ln\bigg( 1 + \frac{\eta\,\delta}{\tau} \bigg) \right] ,
\end{eqnarray}
where $\theta(x)$ is the step function, and 
\begin{equation}\label{etadef}
   \eta = \frac{\tau}{2} \Bigg( \sqrt{1+\frac{4}{\tau}} - 1 \Bigg) .
\end{equation}
Note that the terms inside the square brackets in the expression for 
$f_L$ are 
antisymmetric in $x\leftrightarrow(1-x)$ and thus vanish for the case 
of a symmetric light-cone distribution amplitude.

The above expressions for the distribution functions $w_i(\tau,x)$ 
are a central result of this work. They will allow us to extract
information about renormalon singularities (and hence the structure
of power corrections) as well as about terms of arbitrarily high order
in perturbation theory. Let us now
stress some important general properties of these functions.

\paragraph{IR cancellations:}
Expanding the distribution functions for small $\tau$ shows that 
they vanish at least as fast as $\sqrt\tau$ as $\tau\to 0$, and 
therefore the integrals in (\ref{distr}) are convergent in the IR 
region. As a consequence, the resummed expressions for the 
hard-scattering kernels in (\ref{T1Rfinal}) and (\ref{T1Lfinal}) are 
IR finite. This is a nontrivial result of our analysis, which 
demonstrates the IR finiteness of the hard-scattering kernels (and 
hence factorization) to all orders of perturbation theory 
in the large-$\beta_0$ limit. 
(After \MS subtractions, accomplished by the term proportional 
to $\theta(\tau\,e^{5/3}-1)$ in the expression for $w_L(\tau,x)$, the 
distribution functions vanish like $1/\tau$ for large $\tau$, and hence 
the integrals are also UV convergent.) 

\begin{figure}
\centerline{\epsfxsize=15cm\epsffile{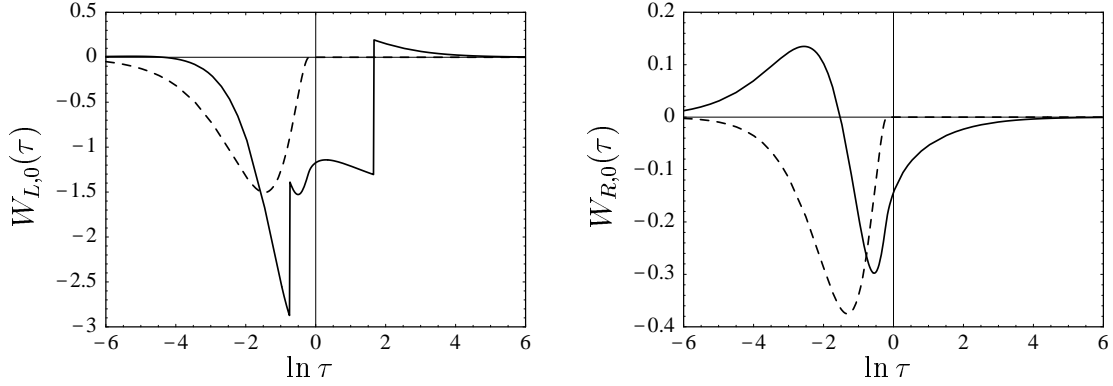}}
\centerline{\parbox{14cm}{\caption{\label{fig:wfuns}
Integrated distribution functions $W_{i,0}(\tau)$ obtained using the 
asymptotic light-cone distribution amplitude. 
Solid (dashed) lines show the real (imaginary) parts.}}}
\end{figure}

\paragraph{Endpoint behavior:}
The proof of IR finiteness of the hard-scattering kernels alone does 
not establish factorization; in addition, one must show that the 
convolutions of the hard-scattering kernels with the light-cone
distribution amplitude of the meson $M$ are convergent. In 
\cite{BBNS}, an explicit calculation showed that at one-loop 
order the hard-scattering kernels tend to a constant (modulo integrable
endpoint logarithms) for $x\to 0$ or $x\to 1$. 
We find that this property 
persists for the resummed perturbation series in the large-$\beta_0$ 
limit. Therefore, the integrals over the light-cone distribution 
amplitude $\Phi_M(x)$ are convergent at the endpoints. (Endpoint 
singularities at least as strong as $1/x^2$ or $1/(1-x)^2$ would be
required to spoil factorization.) We define
\begin{eqnarray}
   W_L(\tau) &=& \int\limits_0^1\!dx\,\Phi_M(x)\,w_L(\tau,x)
    = F_L(\tau,1-z^2) + F_L(\tau/z^2,1-1/z^2) \,, \nonumber\\[-0.2cm]
   W_R(\tau) &=& \int\limits_0^1\!dx\,\Phi_M(x)\,w_R(\tau,x)
    = z\,F_R(\tau,1-z^2) + \frac{1}{z}\,F_R(\tau/z^2,1-1/z^2)
    \,, \quad 
\end{eqnarray}
where the functions $F_i$ are the convolutions of $f_i$ with
the light-cone distribution amplitude. As an example, we perform these
convolutions adopting the asymptotic form $\Phi_0(x)=6x(1-x)$ of 
the light-cone distribution amplitude, valid for light pseudoscalar and
vector mesons in the limit $\mu\to\infty$. This yields
\begin{eqnarray}\label{fints}
   F_{L,0}(\tau,d)
   &=& \frac{3\tau}{2d} \bigg( 7 + \eta + \frac{12\eta}{d} \bigg)
    - \frac{6\tau}{d} \bigg( 1 + \frac{\tau(d+2)}{d^2} 
    + \frac{3\eta}{d} \bigg) \ln\bigg( 1 + \frac{\eta\,d}{\tau} \bigg)
    \nonumber\\
   &&\mbox{}+ \frac{6\tau^2(1+d)}{d^3}\,
    \mbox{Li}_2\bigg(\!-\frac{\eta\,d}{\tau}\bigg)
    + \frac32\,\theta(\tau\,e^{-5/3} - 1) \,, \nonumber\\
   F_{R,0}(\tau,d)
   &=& \frac{3\tau}{2d} \bigg( \!-\! 1 + \eta - \frac{8\eta}{d} \bigg)
    + \frac{6\tau^2}{d^3} \bigg( 1 + \frac{d\eta}{\tau} \bigg)
    \ln\bigg( 1 + \frac{\eta\,d}{\tau} \bigg)
    - \frac{6\tau^2}{d^3}\,
    \mbox{Li}_2\bigg(\!-\frac{\eta\,d}{\tau}\bigg) \,,\qquad
\end{eqnarray}
where $\mbox{Li}_2(x)=-\int_0^x(dt/t)\ln(1-t)$ is the dilogarithm, and
the subscript ``0'' refers to the asymptotic distribution amplitude. In 
Figure~\ref{fig:wfuns}, we show the real and imaginary parts of the 
functions $W_{i,0}(\tau)$ for the case where 
$z=m_c/m_b=0.3$. Note that the steps in the real part of the 
left-handed kernel are an artifact of the \MS subtractions applied to 
the Borel transforms. 

\paragraph{Strong-interaction phases:}
The hard-scattering kernels contain imaginary parts due to gluon 
exchange among the final-state quarks in the third and fourth diagrams 
in Figure~\ref{fig:graphs}. These imaginary parts determine the 
strong-interaction phases of the decay amplitudes in the heavy-quark
limit. They are obtained from the branch cuts of the logarithms in the 
crossed terms in (\ref{wLwR}). Imaginary parts exist for 
$\tau<\delta^2/(z^2+\delta)$ or $\tau<\bar\delta^2/(z^2+\bar\delta)$, 
which in any case implies that $\tau<(1-z^2)^2$.

\paragraph{Perturbative expansion:}
If the running coupling constant under the integral in (\ref{distr}) 
is expanded in powers of a fixed coupling constant using
\begin{equation}\label{alphaexp}
   \alpha_s(\tau\,e^{-5/3}\,m_b^2)
   = \alpha_s(\mu^2) \sum_{n=0}^\infty \bigg( \frac53
   - \log\frac{m_b^2}{\mu^2} - \ln\tau \bigg)^n
   \bigg( \frac{\beta_0\,\alpha_s(\mu)}{4\pi} \bigg)^n \,,
\end{equation}
the integral reproduces the perturbative expansion in powers of 
$\alpha_s(\mu^2)$. Computing the first $n$ terms of the perturbation 
series in the large-$\beta_0$ limit requires evaluating the integrals 
$\int_0^\infty(d\tau/\tau)(\ln\tau)^{n-1} w_i(\tau,x)$ over the 
distribution functions. To perform these integrals analytically, it is 
convenient to change variables from $\tau$ to $\eta$. Below, we derive 
explicit expressions for the hard-scattering kernels at order 
$\beta_0\,\alpha_s^2$. 

The shapes of the distribution functions determine the momentum 
regions from which there arise important contributions to the Borel 
integrals. As a first indicator, one may consider the average value of 
$\ln\tau$, which determines the so-called BLM scale \cite{BLM} through 
$\mu_{\rm BLM}^2=e^{\langle\ln\tau\rangle}\,m_b^2$. (This relation 
holds in the V scheme. In the \MS scheme, the BLM scale 
$\mu_{\rm BLM}$ is reduced by a factor $e^{-5/6}$.) 
As explained in \cite{scale}, a BLM scale is meaningful only in cases 
where the quantity of interest is renormalization-group invariant and 
has a distribution function of definite sign. In our case, this means 
that we can introduce BLM scales only for the imaginary parts of the 
kernels. The dashed curves in Figure~\ref{fig:wfuns} show that these 
scales are significantly smaller than $m_b$. Specifically, we find 
$\mu_{\rm BLM}\approx 0.33 m_b$ ($0.39 m_b$) for the imaginary part of 
the left-handed (right-handed) kernel. When transformed into the \MS 
scheme, the corresponding scales are of order 0.6--0.7\,GeV. This 
suggests that the imaginary parts of the kernels, and hence the 
strong-interaction phases of the decay amplitudes, are susceptible
to nonperturbative physics.

\paragraph{Renormalon ambiguities and power corrections:}
The integrals over the distribution functions in (\ref{distr}) are 
ambiguous because of the Landau pole in the running coupling constant,
\begin{equation}
   \frac{\beta_0\,\alpha_s(\tau\,e^{-5/3} m_b^2)}{4\pi}
   = \frac{1}{\ln\tau + \ln(m_b^2/\Lambda_{\rm V}^2)} \,,
\end{equation}
which is located at $\ln\tau=\ln\tau_L\equiv
\ln(\Lambda_{\rm V}^2/m_b^2)$. The principal values of these integrals 
exactly reproduce the principal values of the original Borel integrals 
\cite{scale}. The residue of the Landau pole therefore provides a 
measure of the renormalon ambiguity, which we define as
\begin{equation}\label{ambi}
   \Delta^{\rm ren}\,[ N_c\,T_{2i}(x) ]
   = \frac{2C_F}{\beta_0}\,w_i(\tau_L,x) \,; \quad
   \tau_L = \frac{\Lambda_{\rm V}^2}{m_b^2} \,.
\end{equation}
Since $\tau_L\ll 1$, the leading contributions to the ambiguity are 
given by the first few terms in the Taylor expansion of the distribution 
functions for small $\tau$.

\section{Large-order behavior of perturbation theory}
\label{sec:pert}

In this section we study in more detail the significance of higher-order
perturbative corrections to the hard-scattering kernels, using the 
all-order results derived in the large-$\beta_0$ limit. We start by 
focusing on the anomalous dimensions of the four-quark operators $O_1$
and $O_2$, whose perturbation series are convergent and can be summed
exactly in the large-$\beta_0$ limit. We then consider higher-order
contributions to the hard-scattering kernels.

\subsection{All-order results for the anomalous dimensions}

Using the explicit result for the function $F_L(\ep,0)$ in 
(\ref{FLep0}), we obtain from (\ref{gamma})
\begin{eqnarray}
   \gamma_\pm &=& \pm \frac{N_c\mp 1}{N_c}\,\frac{b}{\beta_0}\,
    \frac{2(3+2b)(1-b)\,\Gamma(4+2b)}
         {3\Gamma(1-b)\,\Gamma^2(2+b)\,\Gamma(3+b)}
    + O(1/\beta_0^2) \nonumber\\
   &=& \pm \frac{N_c\mp 1}{N_c} \left[ 6\cdot\frac{\alpha_s}{4\pi}
    - \beta_0 \left( \frac{\alpha_s}{4\pi} \right)^2 
    - \frac{65\beta_0^2}{6} \left( \frac{\alpha_s}{4\pi} \right)^3
    + \dots \right] .
\end{eqnarray}
The radius of convergence of the perturbation series is 
$\beta_0|\alpha_s|<4\pi$. The all-order results in the large-$\beta_0$ 
limit may be compared with the exact two-loop expressions \cite{BW92}
\begin{equation}
   \gamma_\pm = \pm \frac{N_c\mp 1}{N_c} \left[
   6\cdot\frac{\alpha_s}{4\pi}
   + \left( -\beta_0 + \frac{N_c}{2} + \frac{57}{2N_c}
   \mp \frac{21}{2} \right) \left( \frac{\alpha_s}{4\pi} \right)^2
   + \dots \right] .
\end{equation}
Numerically, keeping only the term proportional to $\beta_0\,\alpha_s^2$ 
is an excellent approximation for $\gamma_+$ but not for $\gamma_-$. 
However, in both cases the two-loop coefficients are small,
so there is no reason to expect a dominance of the $\beta_0$ terms. 
(These terms are often dominant in cases where the series is
divergent and the expansion coefficients are large.)

\subsection{Partial two-loop results for the hard-scattering kernels}
\label{subsec:2loop}

Perturbative expansions of the hard-scattering kernels can be obtained
by expanding the running coupling constant under the integral in 
(\ref{distr}), as well as the couplings $\alpha_s(m_b^2)$ and 
$\alpha_s(m_c^2)$ in (\ref{T1Lfinal}), in powers of $\alpha_s(\mu^2)$. 
The kernels are then written as
\begin{equation}\label{Tiexpan}
   T_{2i}(x,\mu) = \frac{1}{N_c}\,\delta_{iL} + \frac{2 C_F}{N_c}
   \left[ \frac{\alpha_s(\mu)}{4\pi}\,t_i^{(1)}(x,\mu)
   + \bigg( \frac{\alpha_s(\mu)}{4\pi} \bigg)^2 \beta_0\,t_i^{(2)}(x,\mu)
   + \dots \right] ,
\end{equation}
where $i=L,R$, and $\delta_{iL}=1$ if $i=L$ and zero otherwise. The 
results can be expressed in a compact form by introducing the functions
\begin{eqnarray}
   h_L^{(1)}(\delta,\bar\delta)
   &=& - \frac{3\delta\ln\delta}{2(1-\delta)} + \left[
    \frac{\ln\delta}{1-\delta} - \ln^2\!\delta - \mbox{Li}_2(1-\delta)
    - \{ \delta\to\bar\delta \} \right] , \nonumber\\
   h_R^{(1)}(\delta) &=& - \frac{1}{2(1-\delta)}
    - \frac{\delta\ln\delta}{2(1-\delta)^2} \,, \nonumber\\
   h_L^{(2)}(\delta,\bar\delta)
   &=& \frac{\delta}{4(1-\delta)} \left[ 7\ln\delta 
    - 6\ln^2\!\delta - 6\mbox{Li}_2(1-\delta) \right] \nonumber\\
   &&\mbox{}+ \bigg[ \frac{\ln^2\!\delta - \ln\delta}{1-\delta}
    + \ln(1-\delta) \ln^2\!\delta - \frac23\,\ln^3\!\delta
    + \frac{\mbox{Li}_2(1-\delta)}{1-\delta} \nonumber\\
   &&\quad\mbox{}+ \mbox{Li}_3(1-\delta)
    + 2\mbox{Li}_3(\delta)  - \{ \delta\to\bar\delta \} \bigg] \,,
    \nonumber\\
   h_R^{(2)}(\delta)
   &=& \frac{3}{4(1-\delta)} + \frac{\delta}{4(1-\delta)^2}
    \left[ \ln\delta - 2\ln^2\!\delta - 2\mbox{Li}_2(1-\delta) \right] .
\end{eqnarray}
Here $\mbox{Li}_3(x)=\int_0^x(dt/t)\,\mbox{Li}_2(t)$ is the 
trilogarithm function. At one-loop order we reproduce the expressions 
obtained in \cite{BBNS}, i.e.
\begin{eqnarray}
   t_L^{(1)}(x,\mu) &=& -9 + 6\ln\frac{m_b}{\mu}
    + \left( 3 + 2\ln\frac{x}{1-x} \right) \ln z
    + h_L^{(1)}(\delta,\bar\delta)
    + h_L^{(1)}(-\bar\delta/z^2,-\delta/z^2) \,, \nonumber\\ 
   t_R^{(1)}(x,\mu) &=& z\,h_R^{(1)}(\delta)
    + \frac{1}{z}\,h_R^{(1)}(-\bar\delta/z^2) \,.
\end{eqnarray}
For the corrections of order $\beta_0\,\alpha_s^2$ we find
\begin{eqnarray}
   t_L^{(2)}(x,\mu) &=& -\frac{395}{24} - \pi^2
    - 6\ln^2\!\bigg(\frac{m_b}{\mu}\bigg) + 17\ln\frac{m_b}{\mu}
    \nonumber\\
   &&\mbox{}- \left( 3 + 2\ln\frac{x}{1-x} \right)
    \left( \ln^2\!z + 2\ln z\,\ln\frac{m_b}{\mu} \right)
    + \left( \frac{17}{2} + \frac{10}{3}\,\ln\frac{x}{1-x} \right)
    \ln z \nonumber\\
   &&\mbox{}+ \bigg( \frac53 - 2\ln\frac{m_b}{\mu} \bigg) 
    h_L^{(1)}(\delta,\bar\delta)
    + \bigg( \frac53 - 2\ln\frac{m_b}{\mu} - 2\ln z \bigg)
    h_L^{(1)}(-\bar\delta/z^2,-\delta/z^2) \nonumber\\
   &&\mbox{}- \Big[ h_L^{(2)}(\delta,\bar\delta)
    + h_L^{(2)}(-\bar\delta/z^2,-\delta/z^2) \Big] \,, \nonumber\\ 
   t_R^{(2)}(x,\mu) &=& \bigg( \frac53 - 2\ln\frac{m_b}{\mu} \bigg)
    z\,h_R^{(1)}(\delta)
    + \bigg( \frac53 - 2\ln\frac{m_b}{\mu} - 2\ln z \bigg)
    \frac{1}{z}\,h_R^{(1)}(-\bar\delta/z^2) \nonumber\\
   &&\mbox{}- \Big[ z\,h_R^{(2)}(\delta)
    + \frac{1}{z}\,h_R^{(2)}(-\bar\delta/z^2) \Big] \,.
\end{eqnarray}
The constants $-9$ and $-\frac{395}{24}$ in the expressions for 
$t_L^{(1)}(x,\mu)$ and $t_L^{(2)}(x,\mu)$ are scheme dependent and 
specific to the NDR scheme.

\begin{figure}[t]
\centerline{\epsfxsize=15cm\epsffile{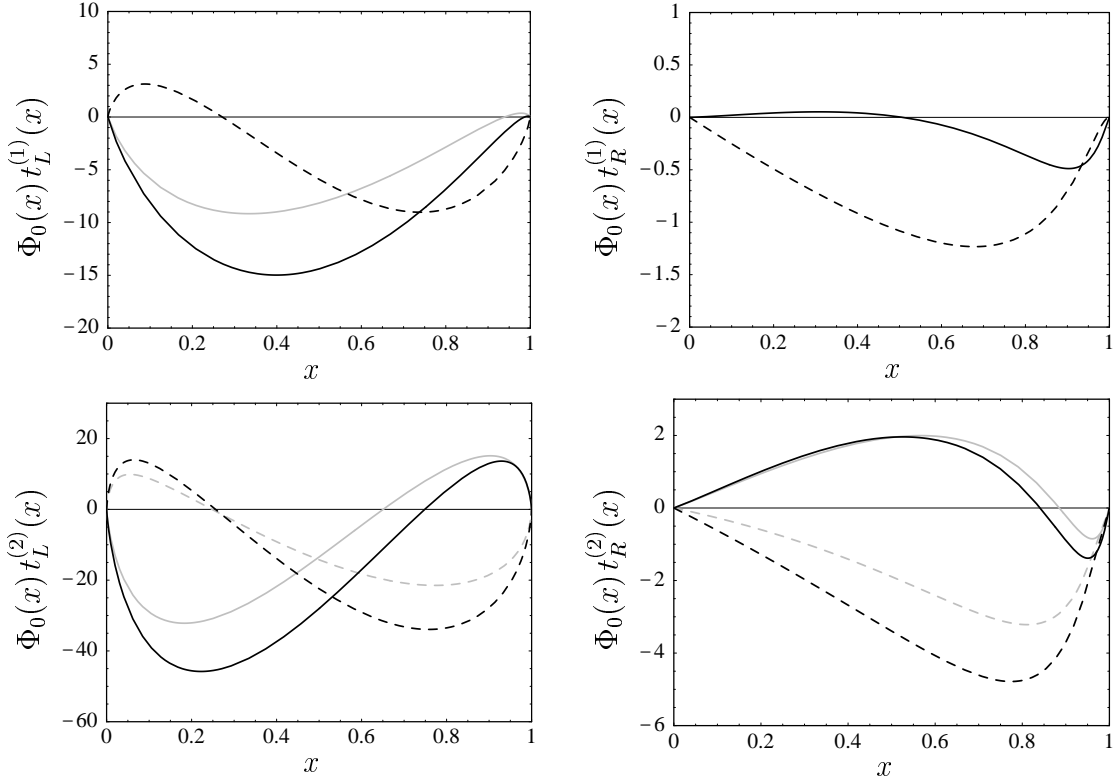}}
\centerline{\parbox{14cm}{\caption{\label{fig:tfuns}
One-loop (upper plots) and partial two-loop (lower plots) contributions 
to the hard-scattering kernels for $\mu=m_b$ (black curves) and, if 
different, $\mu=m_b/2$ (gray curves). Results are multiplied by the 
asymptotic light-cone distribution amplitude. Solid (dashed) lines show 
the real (imaginary) parts.}}}
\end{figure}

In Figure~\ref{fig:tfuns} we show the one-and two-loop kernels for
$z=0.3$ and two values of the renormalization scale. In order to 
suppress the integrable, logarithmic endpoint singularities we have 
multiplied the kernels with the asymptotic distribution amplitude
$\Phi_0(x)=6x(1-x)$. The two-loop contributions 
are sizable. They are suppressed relative to the one-loop contributions 
by a factor $\beta_0\alpha_s/4\pi\approx 0.15$ (for $\mu\approx m_b$), 
but have coefficients that are typically larger by about a factor 3.

\subsection{Numerical results}

The results derived in Section~\ref{sec:Borel} allow us to study the
significance of higher-order perturbative contributions to the 
hard-scattering kernels in the large-$\beta_0$ limit, which may be 
considered as a toy model for the asymptotic behavior of perturbation
theory. In Table~\ref{tab:numerics}, we show the results for the kernels
obtained at fixed order in perturbation theory and compare them
with the principal values of the Borel-resummed series computed from
(\ref{T1Rfinal}), (\ref{T1Lfinal}) and (\ref{distr}). For simplicity, 
the kernels are integrated over $x$ with the asymptotic light-cone 
distribution amplitude. As a measure of the irreducible uncertainty in 
the value of the Borel integral we quote the renormalon ambiguity 
defined in (\ref{ambi}). As input parameters we take $m_b=4.2$\,GeV, 
$z=m_c/m_b=0.3$, and $n_f=4$. We use  the one-loop running coupling 
constant (as is appropriate in the large-$\beta_0$ limit) with 
$\LMS=150$\,MeV, which gives $\alpha_s(m_b^2)=0.226$. The corresponding 
scale parameter in the V scheme is $\Lambda_{\rm V}=345$\,MeV.

\begin{table}[t]
\caption{\label{tab:numerics}
Fixed-order $O(\alpha_s^N)$ perturbative approximations to the 
hard-scattering kernels (convoluted with the asymptotic distribution 
amplitude) and the corresponding principal values of the Borel-resummed 
series, for two choices of the renormalization scale. Results for the 
right-handed kernel are multiplied by 10.}
\begin{center}
\begin{tabular}{|c|cc|cc|}
\hline\hline
 & \multicolumn{2}{|c|}
    {$\int_0^1 dx\,\Phi_0(x)\,N_c\,T_{2L}(x,\mu)$}
 & \multicolumn{2}{|c|}
    {$10\times\int_0^1 dx\,\Phi_0(x)\,N_c\,T_{2R}(x,\mu)$} \\
\hline
$N$ & $\mu=m_b$ & $\mu=m_b/2$ & $\mu=m_b$ & $\mu=m_b/2$ \\
\hline
0 & 1 & 1 & 0 & 0 \\
1 & $0.53-0.18i$ & $0.66-0.23i$
 & $-0.05-0.37i$ & $-0.06-0.47i$ \\
2 & $0.39-0.29i$ & $0.54-0.34i$
 & $\phantom{-}0.02-0.57i$ & $\phantom{-}0.06-0.66i$ \\
3 & $0.34-0.35i$ & $0.52-0.40i$
 & $\phantom{-}0.09-0.68i$ & $\phantom{-}0.15-0.76i$ \\
4 & $0.32-0.40i$ & $0.52-0.45i$
 & $\phantom{-}0.17-0.75i$ & $\phantom{-}0.25-0.81i$ \\
5 & $0.31-0.44i$ & $0.52-0.49i$
 & $\phantom{-}0.24-0.80i$ & $\phantom{-}0.33-0.85i$ \\
6 & $0.31-0.48i$ & $0.52-0.53i$
 & $\phantom{-}0.31-0.83i$ & $\phantom{-}0.43-0.88i$ \\
7 & $0.31-0.51i$ & $0.53-0.58i$
 & $\phantom{-}0.39-0.86i$ & $\phantom{-}0.55-0.90i$ \\
8 & $0.32-0.55i$ & $0.54-0.65i$
 & $\phantom{-}0.49-0.88i$ & $\phantom{-}0.73-0.93i$ \\
9 & $0.33-0.60i$ & $0.58-0.76i$
 & $\phantom{-}0.61-0.91i$ & $\phantom{-}0.99-0.96i$ \\
10 & $0.35-0.67i$ & $0.62-0.94i$
 & $\phantom{-}0.78-0.93i$ & $\phantom{-}1.46-0.99i$ \\
\hline
Borel Sum & $0.29-0.47i$ & $0.49-0.47i$
 & $\phantom{-}0.26-0.90i$ & $\phantom{-}0.26-0.90i$ \\
Ambiguity & $0.003-0.04i$ & $0.003-0.04i$
 & $\phantom{-}0.10-0.03i$ & $\phantom{-}0.10-0.03i$ \\
\hline\hline
\end{tabular}
\end{center}
\end{table}
 
The results in the table show that the large-order behavior of the 
series is improved by lowering the renormalization scale. For 
$\mu=m_b/2$, the two- and three-loop results ($N=2$ or 3) are reasonably 
close to the Borel-resummed value of the entire series. (We stress that 
these resummed values must be considered as estimates only. As a rule of 
thumb, the ``true'' resummed values lie somewhere between the two-loop 
results and the resummed values obtained in the large-$\beta_0$ limit.) 
An exception is the real part of the right-handed kernel, which suffers 
from large cancellations in the integral over the weight function. In 
this case, higher-order contributions in both $\alpha_s$ and $1/\beta_0$ 
are potentially more important. Note that the imaginary parts of the 
kernels approach their asymptotic values slower than the real parts. 
This is a reflection of the low BLM scales obtained earlier. Physically, 
it means that the strong-interaction phases of the decay amplitudes 
are sensitive to nonperturbative hadronization effects.

Another important observation from the table is that the renormalon 
ambiguity, given in the last row, is always much smaller than the
contributions of the first few terms in the perturbation series. (Note 
that the very small ambiguity for the real part of the left-handed 
kernel results from cancellations between various terms in the Taylor 
expansion of the function $W_L(\tau_L)$. The leading power contribution 
alone is 0.017.) The 
series for the kernels are well-behaved in the sense that their 
divergent behavior sets in only in high orders ($N\sim 6$). This 
suggests that power corrections to factorization (at least those
connected with ``nonfactorizable'' soft gluon exchange) are smaller 
than perturbative contributions due to hard gluon exchange. For this 
reason, it would be useful to have complete two-loop expressions
for the hard-scattering kernels. 

As a final remark, we note that for the real part of the left-handed 
kernel the perturbation series never comes close (to within an amount 
of order the renormalon ambiguity) to the principal value of the Borel 
integral. This is because the kernels relevant to
$b\to c\bar u d$ transitions are the sum of two perturbation 
series, one with a ``natural scale'' of order $m_b$ (diagrams 1 and 2),
and the other with a ``natural scale'' of order $m_c$ (diagrams 3 and 
4). These two series reach their minimal term at different values of 
$N$. When their sum is truncated at fixed $N$, one of the two 
series can be close to its minimal term, but not both. This is 
a general feature of multi-scale problems in quantum field theory. If
the scales $m_b$ and $m_c$ were widely separated, one could 
disentangle the two series by constructing an effective field theory.
However, this does not help if, as in the real world, the two scales
are relatively close to each other.

\section{IR renormalons and power corrections}
\label{sec:power}

After the study of higher-order perturbative contributions in the 
previous section, we now turn to the discussion of power-suppressed
effects. We stress, from the outset, that the power 
corrections inferred from the renormalon analysis arise from soft 
``nonfactorizable'' gluon exchange of the type shown in 
Figure~\ref{fig:graphs}. These are not the only sources of
power corrections to the factorization formula (\ref{fact}). For 
instance, different decay topologies such as weak annihilation or 
gluon exchange with the spectator quark in the $B$ meson are known to
contribute at order $\LQCD/m_b$ and have been estimated in \cite{BBNS}. 
For the case of $\bar B^0\to D^{(*)+} M^-$ decays, their effects were 
found to be less than 10\%, as suggested by naive power counting. (For 
the decay $B\to\pi\pi$, the leading power corrections to the 
factorization formula have recently been estimated using light-cone 
QCD sum rules. The result is, again, a moderate correction of less than 
10\% \cite{Khod}.) 

As mentioned earlier, due to the presence of IR renormalon singularities
the resummed perturbation series for the hard-scattering kernels can 
only be defined up to an irreducible ambiguity, which in the 
large-$\beta_0$ limit can be estimated from (\ref{ambi}). We will now
analyze the structure of the leading power-suppressed effects in more
detail, and obtain a minimal model of power corrections that is 
consistent with the renormalon analysis.
It follows from (\ref{ambi}) that the renormalon ambiguity is given in
terms of the distribution functions $w_i(\tau,x)$ evaluated at the 
small value $\tau=\tau_L=(\Lambda_{\rm V}/m_b)^2\ll 1$. Terms of
order $\tau^{k/2}$ in the Taylor expansions of these functions 
correspond to power corrections of order $(\LQCD/m_b)^k$. We focus first 
on the functions
$f_i$ in (\ref{fLRres}) and note that the limits 
$\tau\to 0$ and $x\to 0,1$ do not commute. Naive expansions
of these functions for $\tau\to 0$ would yield more and more powers of
$1/x$ and $1/(1-x)$, thereby introducing endpoint singularities.
However, we have seen earlier that the kernels can be integrated over
$x$ without encountering such singularities. Therefore, the expansions
for small $\tau$ must be written in terms of distributions. For 
simplicity, we shall assume that the light-cone distribution amplitude
$\Phi_M(x)$ vanishes linearly at the endpoints. This is true for the
asymptotic distribution amplitude and, more generally, whenever the
amplitude can be approximated by a finite expansion in Gegenbauer 
polynomials, i.e.
\begin{equation}\label{Gegenb}
   \Phi_M(x,\mu) = 6x(1-x) \bigg[ 1 + \sum_{n\ge 1}^N
   a_n^M(\mu)\,C_n^{(3/2)}(2x-1) \bigg] \,.
\end{equation}
The Gegenbauer moments $a_n^M(\mu)$ are multiplicatively renormalized 
and vanish for $\mu\to\infty$, so that $\Phi_M(x,\mu)$ tends to its 
asymptotic form. Here we adopt the Gegenbauer expansion for
convenience only; our analysis of the endpoint 
behavior in Section~\ref{sec:Borel} showed that the kernels can be 
integrated with any normalizable and smooth function $\Phi_M(x)$.

In our analysis we keep only the leading terms of order $\sqrt\tau$ or 
$\tau$, neglecting higher-order contributions corresponding to power 
corrections of order $(\LQCD/m_b)^3$ and higher. We obtain
\begin{eqnarray}\label{smallt}
   f_L(\tau,\delta,\bar\delta)
   &=& \frac{\tau}{d\,x} \bigg( 2\ln\frac{\sqrt\tau}{d\,x}
    + \frac12 \bigg) \nonumber\\
   &&\mbox{}+ \left[ - \frac{2\sqrt\tau}{d\,x}
    + \frac{\tau}{d^2 x}\,\Delta(x,\sqrt\tau/d) 
    - \{ x\to(1-x) \} \right] + O(\tau^{3/2}) \,, \nonumber\\ 
   f_R(\tau,\delta) &=& - \frac{\tau}{2d\,x} + O(\tau^{3/2}) \,,
\end{eqnarray}
where $d=1-z^2$, and we have introduced the distribution
\begin{equation}
   \Delta(x,a) = \left[ \,\frac{1}{x}
   \bigg( \ln x - \ln a + \frac12 \bigg) \right]_+
   + \frac{\delta(x)}{2} \bigg( \ln^2\!a - \ln a + \frac{\pi^2}{3}
   + \frac12 \bigg) \,.
\end{equation}
Its integral with the light-cone distribution amplitude is defined as
\begin{eqnarray}
   \int\limits_0^1\!dx\,\frac{\Phi_M(x)}{x}\,\Delta(x,a)
   &=& \int\limits_0^1\!dx\,\bigg[ \frac{\Phi_M(x)}{x} - \Phi'_M(0)
    \bigg]\,\frac{1}{x} \bigg( \ln x - \ln a + \frac12 \bigg)
    \nonumber\\
   &&\mbox{}+ \frac{\Phi'_M(0)}{2}\,
    \bigg( \ln^2\!a - \ln a + \frac{\pi^2}{3} + \frac12 \bigg) \,,
\end{eqnarray}
where
\begin{equation}
   \Phi'_M(0) = \lim_{x\to 0} \frac{\Phi_M(x)}{x}
   = 6 \bigg[ 1 + \sum_{n\ge 1} (-1)^n\,\frac{(n+1)(n+2)}{2}\,
   a_n^M(\mu) \bigg] \,.
\end{equation}
The ``+''-distribution for $x\to 1$ is defined in a similar way. Its 
evaluation involves the derivative $\Phi'_M(1)$, which up to an 
overall sign is given by the same sum over Gegenbauer moments, but 
without the factor $(-1)^n$.

To get the corresponding expansions of the functions $w_i(\tau,x)$ we
must add the crossed terms, for which $\tau\to\tau/z^2$ and 
$d\to-d/z^2$. This yields
\begin{eqnarray}
   w_L(\tau,x)
   &=& - \frac{2\tau}{d\,(1-x)}\,(\ln z+i\pi) \nonumber\\
   &&\mbox{}+ \Bigg[ - \frac{2\sqrt\tau}{(1-z)\,x}
    + \frac{\tau}{d^2 x}\,\Big[ \Delta(x,\sqrt\tau/d)
    - z^2\,\Delta(x,-z\sqrt\tau/d) \Big] \nonumber\\
   &&\quad\mbox{}+ \frac{\tau}{d\,x}
    \bigg( 2\ln\frac{\sqrt\tau}{d\,x} + \frac12 \bigg)
    - \{ x\to(1-x) \} \Bigg] + O(\tau^{3/2}) \,, \nonumber\\ 
   w_R(\tau,x) &=& - \frac{\tau}{2d} \bigg( \frac{z}{x} 
    - \frac{1}{z(1-x)} \bigg) + O(\tau^{3/2}) \,.
\end{eqnarray}
The asymmetric part of the function $w_L$ has a small-$\tau$ behavior 
corresponding to a 
first-order power correction in $\LQCD/m_b$, in accordance with the 
result (\ref{pole12}) for the leading renormalon pole. When integrated 
with a symmetric light-cone distribution amplitude this term vanishes,
and the leading power corrections are of order $(\LQCD/m_b)^2$.

In the limit of isospin symmetry, the light-cone distribution amplitude 
for a pion or a $\rho$ meson is symmetric in $x\leftrightarrow(1-x)$. 
We then obtain
\begin{eqnarray}\label{sympower}
   W_L(\tau_L) &=& \frac{6\Lambda_{\rm V}^2}{m_b^2-m_c^2} 
    \bigg( \ln\frac{m_b}{m_c} - i\pi \bigg)
    \bigg( 1 + \!\sum_{n={\rm even}} a_n^M \bigg)
    + O[(\LQCD/m_b)^3] \,, \nonumber\\
   W_R(\tau_L) &=& \frac{3\Lambda_{\rm V}^2}{2m_b m_c} 
    \bigg( 1 + \!\sum_{n={\rm even}} a_n^M \bigg) + O[(\LQCD/m_b)^3] \,,
\end{eqnarray}
where we have used that $\int_0^1(dx/x)\,\Phi_M(x)=3(1+a_2^M+a_4^M+\dots)$
for a symmetric distribution amplitude. For decays involving strange 
particles in the final state, such as $\bar B_0\to D^+\,K^-$, the 
function $\Phi_M(x)$ is no longer expected 
to be symmetric. In such a case, the function $W_L(\tau)$ receives a 
first-order power correction given by 
\begin{equation}\label{antisympower}
   W_L(\tau_L) = \frac{12\Lambda_{\rm V}}{m_b-m_c} \bigg(
   \sum_{n={\rm odd}} a_n^M \bigg) + O[(\LQCD/m_b)^2] \,.
\end{equation}

A comment is in order concerning the peculiar dependence of the power 
corrections in (\ref{sympower}) and (\ref{antisympower}) on the 
heavy-quark masses, which prevents us from taking the limits 
$m_c\to m_b$ or $m_c\to 0$. The limit where the charm-quark mass tends
to zero is actually not a singular one, but has to be dealt with
carefully. We will come back to this in the following section. The fact 
that some power corrections blow up for $m_c\to m_b$ is physical. There 
are three relevant mass scales in this problem: the mass of the 
decaying $b$ quark, the mass of the charm quark, and the energy 
$E_M\approx(m_b^2-m_c^2)/(2 m_b)$ of the light final-state meson. The
factorization properties of the decay amplitude in the heavy-quark 
limit rely crucially on the fact that $E_M\gg\LQCD$, since only then
can the color transparency argument \cite{Bj,DG} be employed to 
demonstrate the cancellation of soft IR contributions \cite{BBNS}. 
This condition no longer holds in the limit $m_c\to m_b$.

Finally, we can use the above results to write down a minimal model for 
the power corrections due to soft gluon exchange in hadronic $B$ decays.
The decay amplitudes for the class-1 decays $\bar B^0\to D^{(*)+} M^-$
are conveniently parameterized in terms of quantities $a_1(D^{(*)} M)$,
which contain the QCD corrections to the results obtained using 
``naive'' factorization. The factorization formula (\ref{fact}) implies 
that
\begin{equation}
   a_1 = \sum_{i=1,2} C_i(\mu) \int\limits_0^1\!dx\,\Phi_M(x,\mu)\,
   \Big[ T_{iL}(x,\mu) \pm T_{iR}(x,\mu) \Big] + O(\LQCD/m_b) \,,
\end{equation}
where the upper (lower) sign refers to the case with a $D$ ($D^*$)
meson in the final state. Inserting the Borel-resummed expressions
for the hard-scattering kernels, we find that the renormalon ambiguity
in $a_1$ is given by
\begin{equation}\label{a1power}
   \Delta^{\rm ren}\,a_1 = \frac{C_2(\mu)}{N_c}\,\frac{2C_F}{\beta_0}\,
   \Big[ W_L(\tau_L) \pm W_R(\tau_L) \Big] + O(1/\beta_0^2) \,.
\end{equation}
To obtain a model for the leading power corrections, we insert here
expressions (\ref{sympower}) or (\ref{antisympower}), depending on 
whether or not the distribution amplitude is symmetric, and replace 
$\Lambda_{\rm V}$ by nonperturbative hadronic parameters 
$\Lambda_{L,R}$. Ideally, these parameters would be determined from a 
fit to experimental data. However, for 
$\Lambda_{L,R}\sim 0.5$\,GeV and a symmetric light-cone distribution
amplitude, we find power corrections of order few times $10^{-3}$, which
are insignificantly small. This is in part due to the 
smallness of the Wilson coefficient $C_2$. The power corrections could
be much larger for other decays, such as $B\to\pi\pi$. For
decays with a strange meson in the final state, such as 
$\bar B^0\to D^+ K^-$, our model predicts first-order power corrections 
as large as a few percent. Unfortunately, these modes are 
Cabibbo suppressed, and no experimental data has been reported yet.

\section{Results for charmless hadronic decays}
\label{sec:zto0}

Our discussion so far has referred to the simplest application of the 
QCD
factorization formula, i.e.\ to class-1 $B$ decays into a heavy--light
final state. However, factorization in the heavy-quark limit also occurs
for the phenomenologically more interesting, rare hadronic decays into
two light mesons. Examples are the charmless decays $B\to\pi\pi$ and
$B\to\pi K$, which might provide information about
the CP-violating phase of the quark mixing matrix \cite{BBNS1}. The
effective weak Hamiltonian for these processes contains many penguin
operators besides the current--current operators $O_1$ and $O_2$, and
several different decay topologies must be considered in addition to 
the diagrams shown in Figure~\ref{fig:graphs}. A complete renormalon 
analysis for these processes is beyond the scope of this paper; however, 
the ``nonfactorizable'' vertex corrections investigated here are an
important part of such an 
analysis. It is thus interesting to apply our results to the case
where the charm quark in the final state is replaced by a massless 
$u$ quark.

The limit $z\to 0$ is smooth but must be taken carefully. For instance, 
the power corrections in (\ref{sympower}) cannot be extrapolated to 
$m_c\to 0$, but we obtain regular expressions by first computing the 
distribution functions in the limit $z\to 0$, and then 
expanding the results for small $\tau$. We now collect the most 
important formulae valid for $z=0$.

\subsection{Borel representation and distribution function}

In the massless limit the chirality of the external quark states is 
preserved, and hence the right-handed kernel $T_R$ vanishes for $z\to 0$.
This can also be seen explicitly by taking the limit $z\to 0$ in our
results for this kernel. The Borel-resummed expression (\ref{T1Lfinal})
for the left-handed kernel simplifies to
\begin{eqnarray}
   N_c\,T_{2L}(x,\mu) &=& 1 + \frac{2 C_F}{\beta_0} \Bigg(
    \int\limits_{\,-b(\mu)}^0 \frac{d\ep}{\ep}\,\left[ 
    3 - \frac{(3-2\ep)(1+\ep)\Gamma(4-2\ep)}
             {3\Gamma(1+\ep)\,\Gamma^2(2-\ep)\,\Gamma(3-\ep)} \right]
    + 3\,\ln\frac{b(\mu)}{b(m_b)} \Bigg) \nonumber\\
   &&+ 2 C_F \int\limits_0^\infty\frac{d\tau}{\tau}\,w_L(\tau,x)\,
    \frac{\alpha_s(\tau\,e^{-5/3}\,m_b^2)}{4\pi} + O(1/\beta_0^2) \,,
\end{eqnarray}
where
\begin{eqnarray}\label{wLlight}
   w_L(\tau,x) 
   &=& - \tau \bigg( \frac{1-\eta}{2x} + \frac{\eta}{x^2} \bigg)
    - \frac{\tau\,[2x^2-\tau(1-x)]}{x^3}\,
    \ln\bigg( 1 + \frac{\eta\,x}{\tau} \bigg) \nonumber\\
   &&\mbox{}+ \frac{\tau(2x-\tau)}{x^2}\,
    \ln\bigg( 1 - \frac{x}{\tau} - i\epsilon \bigg) 
    + 3\,\theta(\tau\,e^{-5/3}-1) \nonumber\\
   &&\mbox{}+ \Bigg[ \,\frac{\tau-\eta}{x}
    - \bigg( 1 - \frac{\tau}{x^2} \bigg) 
    \ln\bigg( 1 + \frac{\eta\,x}{\tau} \bigg) 
    + \bigg( 1 - \frac{\tau}{x} \bigg)^2 
    \ln\bigg( 1 - \frac{x}{\tau} - i\epsilon \bigg) 
    \nonumber\\
   &&\quad\mbox{} - \{x\to(1-x)\} \Bigg] .
\end{eqnarray}

\begin{figure}
\centerline{\epsfxsize=15cm\epsffile{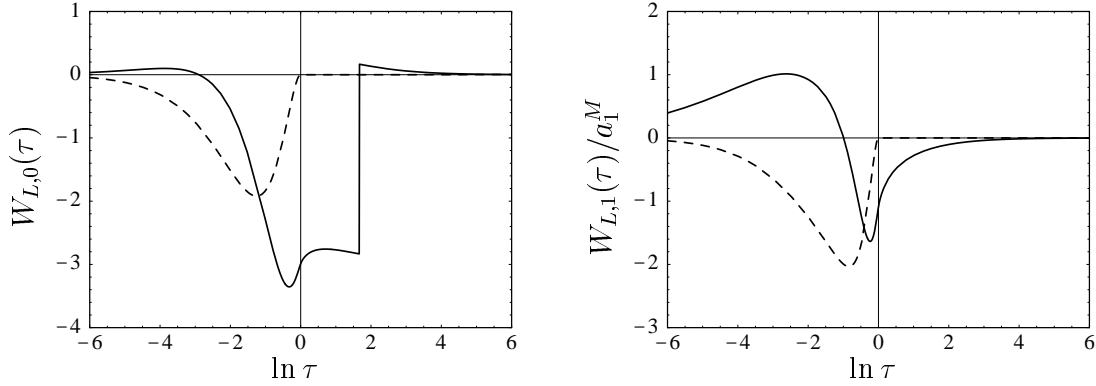}}
\centerline{\parbox{14cm}{\caption{\label{fig:wLlight}
Integrated distribution function $W_L(\tau)$ for $m_c=0$. The left-hand 
plot shows the asymptotic result, and the right-hand one the 
contribution proportional to the first Gegenbauer moment.}}}
\end{figure}

As before, the kernel is well-behaved in the endpoint regions $x\to 0$ 
and $x\to 1$. In particular, the integral of $w_L(\tau,x)$ with the 
asymptotic light-cone distribution amplitude yields
\begin{eqnarray}
   W_{L,0}(\tau) 
   &=& - \frac{3\tau}{2}\,(1-13\eta)
    - 6\tau(1+3\eta+3\tau)\,\ln\bigg( 1 + \frac{\eta}{\tau} \bigg)
    + 6\tau(1-\tau)\,\ln\bigg( 1 - \frac{1}{\tau} - i\epsilon \bigg)
    \nonumber\\
   &&\mbox{}+ 12\tau^2\,\mbox{Li}_2\bigg( \!-\frac{\eta}{\tau} \bigg)
    + 6\tau^2\,\mbox{Li}_2\bigg( \frac{1}{\tau} + i\epsilon \bigg)
    + 3\,\theta(\tau\,e^{-5/3}-1) \,.
\end{eqnarray}
This function is shown in the left-hand plot in 
Figure~\ref{fig:wLlight}. In the second plot, we show the contribution
proportional to the first Gegenbauer moment, denoted by $W_{L,1}(\tau)$. 
It is evident that the real part of $W_{L,1}(\tau)$ receives large 
contributions from the region of very small $\tau$, corresponding to 
low gluon virtualities. This is in accordance with the fact that the
renormalon ambiguity for this contribution is of first order in
$\LQCD/m_b$. 

\subsection{Partial two-loop results for the hard-scattering kernel}

The easiest way to obtain the one and two-loop coefficients of the
kernel in the massless case is to take the limit $z\to 0$ in the 
expressions collected in Section~\ref{subsec:2loop}. It is a nontrivial
check of our results that all logarithms of the mass ratio $z$ cancel
in that limit. For the expansion coefficients in (\ref{Tiexpan}) we 
obtain
\begin{eqnarray}
   t_L^{(1)}(x,\mu) &=& -9 + 6\ln\frac{m_b}{\mu} + \frac32 \bigg(
    \frac{1-2x}{1-x}\,\ln x - i\pi \bigg) \nonumber\\
   &&\mbox{}+ \left[ - \frac{\ln^2\!x}{2} + \frac{\ln x}{1-x}
    + \mbox{Li}_2(x) - \bigg( \frac32 + i\pi \bigg) \ln x
    - \{ x\to(1-x) \} \right] , \nonumber\\
   t_L^{(2)}(x,\mu) &=& 6\ln^2\!\bigg( \frac{m_b}{\mu} \bigg)
    - 11\ln\frac{m_b}{\mu} - \frac{35}{24} 
    + \bigg( \frac53 - 2\ln\frac{m_b}{\mu} \bigg)\,t_L^{(1)}(x,\mu)
    \nonumber\\
   &&\mbox{}- \frac{3(1-3x)}{4(1-x)}\,\ln^2\!x
    + \frac{7(1-2x)}{4(1-x)}\,\ln x
    + \frac{3(1-x)}{2x}\,\mbox{Li}_2(x)
    + i\pi \bigg( \frac32\,\ln x - \frac74 \bigg) \nonumber\\
   &&\mbox{}+ \Bigg[ \,\frac{\ln^3\!x}{2} - \ln^2\!x \ln(1-x)
    - \frac{1+3x}{4(1-x)}\,\ln^2\!x + \bigg( \pi^2 - \frac74
    + \frac{1}{1-x} \bigg) \ln x \nonumber\\
   &&\quad\mbox{}- \frac{1-3x}{2x}\,\mbox{Li}_2(x) - \mbox{Li}_3(x)
    + \frac{i\pi}{2}\,(\ln^2\!x - 3 \ln x) - \{ x\to 1-x \} \Bigg] .
\end{eqnarray}

\subsection{Numerical results}

In Table~\ref{tab:num}, we show the results for the kernel obtained at 
fixed order in perturbation theory and compare them with the 
principal value of the Borel-resummed series. To illustrate 
the effect of the leading IR renormalon singularity, we include 
the contribution of the first Gegenbauer moment $a_1^M$. Then the 
light-cone distribution amplitude is no longer symmetric, and the 
leading renormalon ambiguity is of first order in $\LQCD/m_b$. 

\begin{table}[t]
\caption{\label{tab:num}
Fixed-order $O(\alpha_s^N)$ perturbative approximations to the 
hard-scattering kernel in the limit $m_c\to 0$ and the corresponding 
principal value of the Borel-resummed series, for two 
choices of the renormalization scale. Shown are the first two terms in 
the Gegenbauer expansion of the distribution amplitude.}
\begin{center}
\begin{tabular}{|c|c|c|}
\hline\hline
 & \multicolumn{2}{|c|}
    {$\int_0^1 dx\,\Phi_M(x,\mu)\,N_c\,T_{2L}(x,\mu)$} \\
\hline
$N$ & $\mu=m_b$ & $\mu=m_b/2$ \\
\hline
0 & 1 & 1 \\
1 & $(0.56-0.23i)+a_1^M (0.13-0.23i)$
  & $(0.69-0.29i)+a_1^M (0.17-0.29i)$ \\
2 & $(0.44-0.35i)+a_1^M (0.31-0.35i)$
  & $(0.61-0.41i)+a_1^M (0.41-0.40i)$ \\
3 & $(0.41-0.43i)+a_1^M (0.51-0.42i)$
  & $(0.61-0.48i)+a_1^M (0.67-0.46i)$ \\
4 & $(0.41-0.48i)+a_1^M (0.77-0.46i)$
  & $(0.62-0.53i)+a_1^M (1.06-0.51i)$ \\
5 & $(0.42-0.52i)+a_1^M (1.13-0.50i)$
  & $(0.65-0.57i)+a_1^M (1.71-0.54i)$ \\
6 & $(0.44-0.55i)+a_1^M (1.76-0.53i)$
  & $(0.67-0.61i)+a_1^M (3.05-0.58i)$ \\
7 & $(0.46-0.59i)+a_1^M (2.96-0.57i)$
  & $(0.72-0.66i)+a_1^M (6.28-0.63i)$ \\
8 & $(0.49-0.63i)+a_1^M (5.63-0.60i)$
  & $(0.78-0.72i)+a_1^M (15.2-0.69i)$ \\
9 & $(0.54-0.68i)+a_1^M (12.3-0.65i)$
  & $(0.91-0.82i)+a_1^M (42.7-0.79i)$ \\
10 & $(0.61-0.74i)+a_1^M (30.5-0.71i)$
  & $(1.15-0.99i)+a_1^M (138.-0.96i)$ \\
\hline
Borel Sum & $(0.38-0.55i)+a_1^M (0.27-0.53i)$
 & $(0.59-0.55i)+a_1^M (0.27-0.53i)$ \\
Ambiguity & $(0.02-0.04i)+a_1^M (0.19-0.04i)$
 & $(0.02-0.04i)+a_1^M (0.19-0.04i)$ \\
\hline\hline
\end{tabular}
\end{center}
\end{table}

The results obtained using the asymptotic distribution amplitude
exhibit a similar behavior as in the case with 
a finite charm-quark mass. The convergence of the series is improved
by using the smaller renormalization scale $\mu=m_b/2$. In this case,
the two-loop result is already very close to the asymptotic value. As
previously, the imaginary part of the kernel reaches its asymptotic 
value at larger $N$ than the real part.
The situation for the contribution proportional to the first Gegenbauer 
moment is different. Whereas the imaginary part shows a similar
behavior as for the leading term, the series for the real part of 
$W_{L,1}(\tau)$ diverges already in low orders.
Even the two-loop results exceed the asymptotic values, and
starting at $N\sim 4$ the expansion coefficients exhibit 
a rapid factorial
growth. This is a reflection of the leading renormalon pole at 
$u=\frac12$, corresponding to a first-order power correction to the real
part, which is
absent in the case of a symmetric distribution amplitude. Accordingly,
the renormalon ambiguity is an order of magnitude larger than in the 
symmetric case.

Finally, we note that using the one-loop expressions for the kernel, as
is done in all phenomenological applications of the QCD factorization
approach to date, gives a reasonable approximation to the real part
(for $\mu=m_b/2$), but underestimates the strong-interaction phase by
almost a factor 2. This observation may be relevant for studies of CP
asymmetries in rare hadronic $B$ decays.

\subsection{IR renormalons and power corrections}

The pattern of IR renormalons in the massless limit is very similar to
that for finite charm-quark mass. Following our analysis in
Section~\ref{sec:power}, we first construct the expansion of the 
distribution function (\ref{wLlight}) for small values of $\tau$. The
result can be written as
\begin{eqnarray}
   w_L(\tau,x) 
   &=& - \frac{\tau}{1-x} \bigg( \ln\tau + \frac12 + 2i\pi \bigg)
    \\
   &&\mbox{}+ \left[ - \frac{2\sqrt\tau}{x} + \frac{\tau}{x} \bigg(
    2\ln\frac{\sqrt\tau}{x} - \frac12 \bigg) 
    + \frac{\tau}{x}\,\widetilde\Delta(x,\tau) - \{ x\to(1-x)\}
    \right] + O(\tau^{3/2}) \,, \nonumber
\end{eqnarray}
where
\begin{equation}
   \widetilde\Delta(x,\tau) = \left[ \,\frac{1}{x} \bigg( \ln x
   - \frac12\,\ln\tau + \frac12 \bigg) \right]_+
   + \delta(x) \bigg( \,\frac18\,\ln^2\!\tau
   - \frac14\,\ln\tau + \frac{\pi^2}{6} + \frac14 \bigg) \,.
\end{equation} 
After integration with the light-cone distribution amplitude, we find 
that for the case of a symmetric amplitude the leading power-suppressed 
contributions are given by
\begin{equation}
   W_L(\tau_L) = \frac{6\Lambda_{\rm V}^2}{m_b^2} 
    \bigg( \ln\frac{m_b}{\Lambda_{\rm V}} - \frac14 - i\pi \bigg)
    \bigg( 1 + \!\sum_{n={\rm even}} a_n \bigg) + O[(\LQCD/m_b)^3] \,,
\end{equation}
which may be compared with the corresponding result in (\ref{sympower}). 
If the distribution amplitude is not symmetric, the function $W_L(\tau)$ 
receives a first-order power correction given by 
\begin{equation}
   W_L(\tau_L) = \frac{12\Lambda_{\rm V}}{m_b} \bigg(
   \sum_{n={\rm odd}} a_n \bigg) + O[(\LQCD/m_b)^2] \,.
\end{equation}

As in Section~\ref{sec:power}, these results could be used to construct 
a simple model of power corrections to factorization in rare hadronic 
$B$ decays. However, as mentioned earlier, this model would be
incomplete without taking into account other decay topologies such as
penguin contractions and interactions with the 
spectator quark in the $B$ meson. We plan to analyze these contributions 
in a future publication.

\section{Conclusions}
\label{sec:concl}

We have used the renormalon calculus to study the asymptotic behavior
of the hard-scattering kernels entering the QCD factorization formula 
for the nonleptonic weak decays $\bar B^0\to D^{(*)+} M^-$. We have
obtained explicit results for the Borel transforms and momentum 
distribution functions of the kernels in the approximation of retaining 
a single renormalon chain (the large-$\beta_0$ limit). This method 
estimates power corrections to the factorization formula, and allows 
us to investigate the (divergent) higher-order behavior of the 
perturbation series for the kernels. From the pattern of singularities
in the Borel plane, we have derived a simple model of power corrections 
that is consistent with the renormalon analysis. This model accounts 
only for corrections due to soft, ``nonfactorizable'' 
gluon exchange. Other types of power-suppressed effects exist and
have been estimated in the literature \cite{BBNS}.

An unexpected result of our work is that the leading IR renormalon
singularity, corresponding to a first-order power correction in
$\LQCD/m_b$, vanishes for mesons with a symmetric light-cone 
distribution amplitude. We have explicitly calculated the second-order 
correction and shown it to be numerically small. Higher-order
perturbative effects are thus expected to be more 
important than power corrections. We have presented analytic results
for the contributions of order $\beta_0\alpha_s^2$ to the 
hard-scattering kernels, which presumably are the dominant part of the
full two-loop contributions. We have also given numerical results for 
the terms of order $\beta_0^{n-1}\alpha_s^n$ in the perturbation series.
We have shown that lowering the renormalization scale below $m_b$ 
improves the rate of approach of the series to their asymptotic values. 
The BLM scales associated with the 
imaginary parts of the hard-scattering kernels are below 1\,GeV (in
the \MS scheme). This indicates that the strong-interaction phases of 
the decay amplitudes are not insensitive to nonperturbative physics.
Finally, by showing that the kernels are free of IR divergences and
power-divergent endpoint singularities, we have proven the factorization 
formula presented in \cite{BBNS} to all orders in perturbation theory
in the large-$\beta_0$ limit.

In the future, it would be worthwhile to carry out a renormalon 
analysis for the phenomenologically more interesting charmless hadronic 
$B$ decays. Power corrections in 
these processes are, in general, not suppressed by small Wilson 
coefficient functions. By giving explicit results valid in the limit 
$m_c\to 0$, we have accomplished the technically most challenging part 
of such an analysis. It remains to add the contributions from penguin 
contractions and gluon exchange with the spectator quark.

\vspace{0.3cm}
{\it Acknowledgements:\/}
We are grateful to Gerhard Buchalla for helpful discussions. This work 
was supported in part by the National Science Foundation. T.B.~is 
supported by the Swiss National Science Foundation. 

\vspace{0.3cm}
{\it Note added:\/}
While this paper was in writing, the work {\em Renormalon analysis of 
heavy--light exclusive B decays\/} appeared \cite{Burrell}, in which
the authors present expressions for the Borel transforms of the 
hard-scattering kernels that are equivalent to our eq.~(\ref{Fi0u}), 
and observe the vanishing of first-order power corrections for the case 
of a symmetric light-cone distribution amplitude, in accordance with 
our eq.~(\ref{pole12}).

\end{document}